\documentclass{emulateapj}

\def\RSUN{R$_{\sun}~$}

\def\C3{$\rm C~III$}
\def\OO5{$\rm O~V$}
\def\N3{$\rm N~III$ }
\def\O6{$\rm O~VI$}
\def\fe18{$\rm [Fe~XVIII]$}
\def\Si12{$\rm Si~XII$}
\def\Al11{$\rm Al~XI$}
\def\si8{$\rm [Si~VIII]$}
\def\Fe10{$\rm [Fe~X]$}
\def\Ni14{$\rm [Ni~XIV]$}
\def\Ca14{$\rm [Ca~XIV]$}
\def\SSi3{$\rm Si~III$ }

\begin{document}

\shorttitle{Spatial Offsets in Flare-CME Current Sheets}
\shortauthors{Raymond, Giordano \& Ciaravella}

\title{Spatial Offsets in Flare-CME Current Sheets}

\author{John C. Raymond,}
\affil{Harvard-Smithsonian Center for Astrophysics, 60 Garden St.,
Cambridge, MA  02138, USA}
\email{jraymond@cfa.harvard.edu}

\author{Silvio Giordano}
\affil{INAF-Osservatorio Astrofisico di Torino, via Osservatorio 20, 
I-10025 Pino Torinese, Italy}

   \and

\author{Angela Ciaravella}
\affil{INAF-Osservatorio Astronomico di Palermo, P.za Parlamento 1, I-90134 Palermo, Italy}

\begin{abstract}
Magnetic reconnection plays an integral part in nearly all models
of solar flares and coronal mass ejections (CMEs).   The reconnection heats
and accelerates the plasma, produces energetic electrons and ions, and 
changes the magnetic topology to form magnetic flux ropes and allow CMEs to
escape.  Structures that appear between flare loops and CME cores in optical, 
UV, EUV and X-ray observations have been identified as current sheets and
interpreted in terms of the nature of the reconnection process and the energetics
of the events.  Many of these studies have used UV spectral observations of high
temperature emission features in the [Fe~XVIII] and Si~XII lines.  In this paper
we discuss several surprising cases in which the [Fe~XVIII] and Si~XII emission peaks are
spatially offset from each other.  We discuss interpretations based on asymmetric reconnection,
on a thin reconnection region within a broader streamer-like structure, and
on projection effects.  Some events seem to be easily interpreted as
projection of a sheet that is extended along the line of sight that is viewed
an angle, but a physical interpretation in terms of asymmetric reconnection is
also plausible.  Other events favor an interpretation as a thin current sheet
embedded in a streamer-like structure.

\end{abstract}


\section{Introduction}

Most large solar eruptions involve both an X-ray flare and a coronal mass
ejection.  While different magnetic topologies and trigger 
mechanisms are invoked in different scenarios, essentially all of the models
include a current sheet where magnetic field rapidly reconnects.  The reconnection
converts magnetic free energy into heat, kinetic energy and energetic particles to
power the flare, and it changes the magnetic topology, which helps the CME to escape.
The reconnection also produces a twisted topology, either wrapping more helical
field around an existing flux rope \citep{linforbes, lin04} or creating a flux rope
from scratch \citep{gosling95}.

A common feature of all flare models is that while magnetic reconnection must be
slow under normal circumstances to allow magnetic free energy to build up in the
corona before the eruption, fast reconnection is required to match the rapid rise of flare emission
during the eruption.
In most models reconnection must occur over large area to account for the total energy.  On
small scales, kinetic effects and the tearing mode instability are crucial 
\citep{loureiro, jidaughton, shen13a},
while on larger scales the current sheet is described either as a small diffusion
region at an X-line with larger scale shock-like structures enclosing an exhaust region 
(Petschek model) or as a thick turbulent current sheet \citep{lazarianvishniac, kowal09}.

Various observations of flares and CMEs have been interpreted in terms of current sheets
and used to estimate such parameters as the temperature, thickness, density, turbulent velocity
and reconnection rate.  The Ultraviolet Coronagraph Spectrometer (UVCS) \citep{kohl95, kohl97} 
observed the high temperature emission line of [Fe XVIII] in narrow
features between the post-flare loops and CME cores \citep{ciaravella02, ko03, bemporad06, ciaravellaraymond,
schettino10}, and even higher temperature lines were observed with Solar Ultraviolet
Measurements of Emitted Radiation (SUMER) experiment
\citep{innes03a, innes03b} and the Interface Region Imaging Spectrograph (IRIS) \citep{tian14}.  
Narrow band EUV images from Extreme-Ultraviolet Imaging
Telescope (EIT) and TRACE \citep{yokoyama, mckenzie}
and the Atmospheric Imaging Assembly (AIA) \citep{reevesgolub, savage12} have 
been interpreted in terms of inflow,
outflow and heating in the reconnection region.  White light features that resemble rays
topped by disconnection regions seen in the wakes of some CMEs seem to be current sheets
in many cases \citep{webb03, ciaravella13}, and hard X-ray emission seems to show hot regions
on either side of an X-line in some RHESSI observations \citep{suiholman, susino13}

In the course of comparing white light and UVCS observations of candidate current sheets,
\cite{ciaravella13} compiled a list of events detected in [Fe~XVIII] by UVCS.  The [Fe~XVIII]
$\lambda$975 line emission peaks at log T=6.85 according to version 7 of CHIANTI \citep{landi13}, and
it is rarely seen with UVCS except in CMEs and current sheets.  Several of
the events in \cite{ciaravella13} showed a significant offset between the peak spatial positions of [Fe~XVIII] and
Si~XII $\lambda$$\lambda$499, 521.  In this paper we quantify those offsets and discuss interpretations in terms of
actual physical offsets due to asymmetric reconnection or of projection effects resulting
from viewing geometry.  Projection effects are a possible explanation for the events with
a simple offset, but are less likely to account for events where the [Fe~XVIII] emission
is centered within a wider Si~XII feature.  The observations are described in Section 2, while the analysis 
and results in Section 3. In Section 4 we present the interpretation of the CS characteristics. The discussion 
and conclusions are in Section 5.

\section{Observations}

We examined the events from the \cite{ciaravella13} list of 28 current sheets detected in the
[Fe~XVIII] line with UVCS.  We excluded events not detected in Si~XII and events in which
there were too few counts in the [Fe~XVIII] line to obtain a good intensity profile along the
UVCS slit.  The remaining 14 events 
are listed in Table 1,
and details of several individual events are discussed in the Appendix.  
The number of events is modest because the [Fe~XVIII] line was only observable in a small 
range of grating positions, and the UVCS instrument configuration only covered that wavelength
for a small fraction of the time that UVCS was observing.  However, 
almost all UVCS observations covered the O~VI doublet, at least one member of the Si~XII
doublet, and either Ly$\alpha$ or Ly$\beta$. The emissivities of these lines peak at 
different temperatures; log T = 6.30 for Si~XII and log T = 5.50 for O~VI, but both have 
significant emissivity at higher temperatures.  The Lyman line emission drops off 
approximately as $T^{-0.5}$, but it also depends on the outflow speed and temperature of the plasma 
through Doppler dimming.  

The UVCS instrument measures the intensities and line profiles of coronal emission lines 
along a 42$^\prime$ slit which is perpendicular to the radial vector from Sun center to a reference
point near the middle of the slit. The slit could be placed at any heliocentric height, but because the 
intensity drops off rapidly with distance above the surface and we are studying faint
lines, all the observations reported here were at 1.91 \RSUN from Sun center or lower.
The spatial resolution was generally
limited by onboard binning mandated by the telemetry rate, and it ranges from 
21$^\prime$$^\prime$ to 70$^\prime$$^\prime$ for the observations presented here.
Many different observing sequences were employed during
the life of the mission, so the exposure times vary from a few minutes to a few hours.
More details about the observations are given in \cite{ciaravella13} and \cite{giordano13}.

To investigate the spatial relationships among the emission at different temperatures,
we measured the intensity profiles of the different spectral lines along the slit.  We 
measured the intensities in two ways in order to assess uncertainties in background
subtraction:  First, we simply added up the counts above background in each spatial bin, 
and second, we fit a Gaussian
to the profile in each spatial bin.  The results for the two measurements are nearly
indistinguishable except in cases where there are too few counts for a meaningful Gaussian
fit, and in those cases the intensities from adding up the counts were too noisy to
produce significant correlations among the line intensities.

For each event (and for each slit position if more than one were obtained) we performed a cross
correlation between the spatial profiles of [Fe~XVIII] and the Si~XII, O~VI and Lyman
lines in order to quantify the separations between features seen at different temperatures in an
objective manner.   
The spatial morphology of [Fe~XVIII] intensity is a single or double peak. In the latter cases the correlation analysis has been performed for each peak separately. The results will be discussed in more detail in section~\ref{res}.

\section{Analysis and Results}\label{res}

The results of our cross correlation analysis are  in Table~\ref{tab}. 
For each event we list  the data and time of the observation, the height, Position Angle, the spatial width of the [Fe~XVIII] line and, the cross correlation parameters (factor, lag and width) with Si~XII, O~VI and   Ly$\beta$($\alpha$) lines. 
Here, the lag is the spatial separation between the peaks seen in different lines. 
The spatial width has been computed as the FWHM of the  [Fe~XVIII] intensity distribution along the UVCS slit.  In the last column of Table 1, we also give a classification based on the spatial profiles of [Fe~XVIII] and Si~XII;  "S" if the profiles are
basically the same, "O" if they are similar but offset from each other, and "I" if the [Fe~XVIII]
emission lies within a broader Si~XII feature.
The 3 events  marked with 'b' in Table~\ref{tab} are those in which [Fe~XVIII] spatial profiles show two  peaks. 
Generally the two peaks  do not appear in the UV spectra simultaneously. In some cases one peak appears during the observation  as associated to 
a CME. In other cases the peaks are detected since the beginning of the observation, and they may or may not be easily associated with a CME observed in WL.
The two peaks are often different in intensity and spatial distribution. 
We analyse them separately and in Table~\ref{tab} we present the results for the peak that has been associated with a CME in WL. 

\begin{figure*}[h]
\center
   \begin{tabular}{ccc}
	\includegraphics[width=5.5cm]{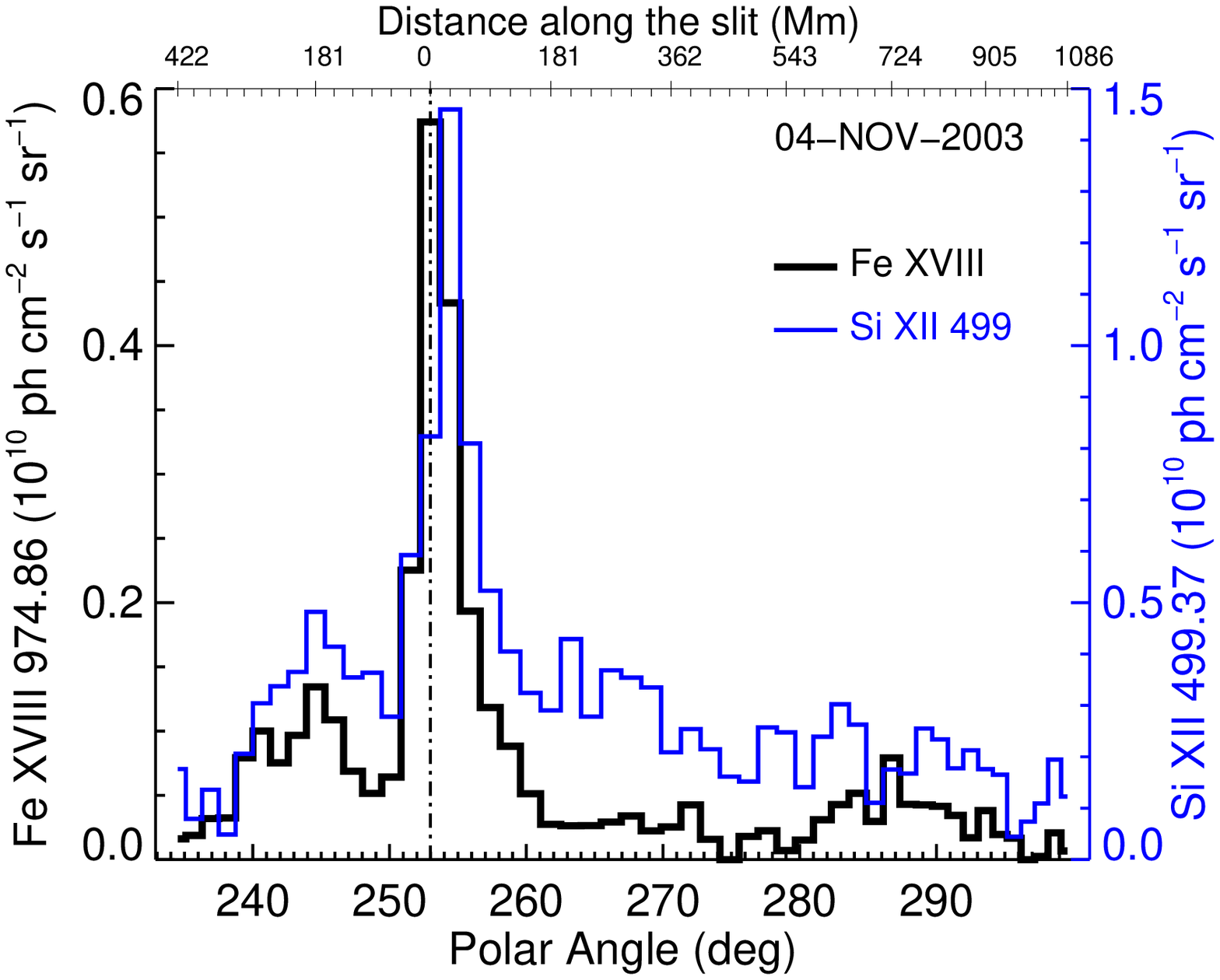} & 	
	\includegraphics[width=5.5cm]{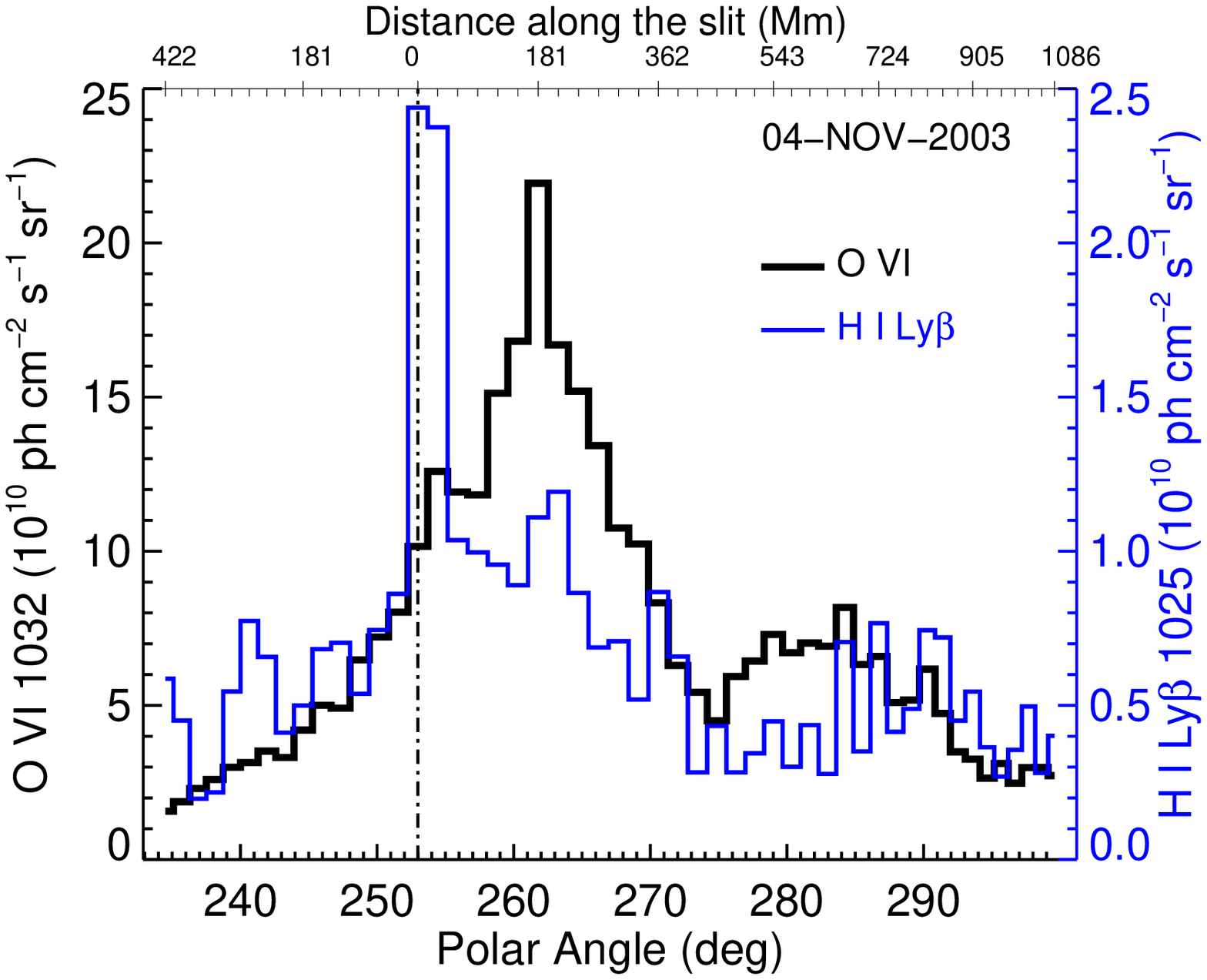} &
 	\includegraphics[width=5.5cm]{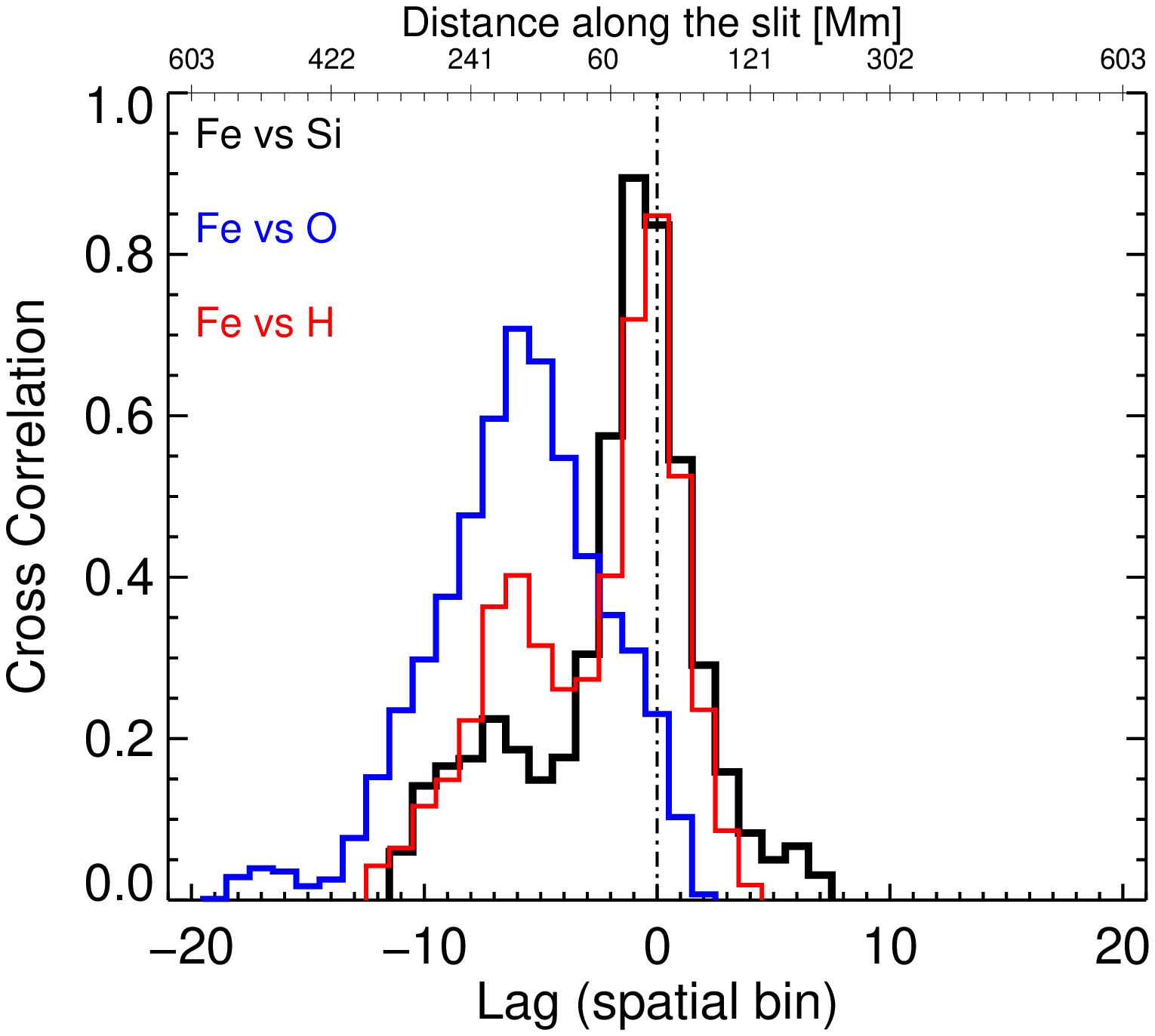} \\
 \end{tabular}
  \caption{The event of 2003 November 4 as an example of narrow, coincident features
in Si~XII and [Fe~XVIII] (class "S").  The left panel shows the
intensities of the [Fe~XVIII] and Si~XII lines (with the Si~XII scale on the right)
as functions of position along the slit with UVCS mirror pointed to 1.67~\RSUN. The vertical dashed line indicates the peak of the [Fe~XVIII]
intensity.  In the middle panel are the O VI $\lambda$1032 and Ly$\beta$ lines.  The right panel shows 
the cross correlations between [Fe~XVIII] and Si~XII (black),
O VI (blue) and Ly$\alpha$ (red).}
\label{fig1}
\end{figure*}

\begin{figure*}[h]
\center
   \begin{tabular}{ccc}
	\includegraphics[width=5.5cm]{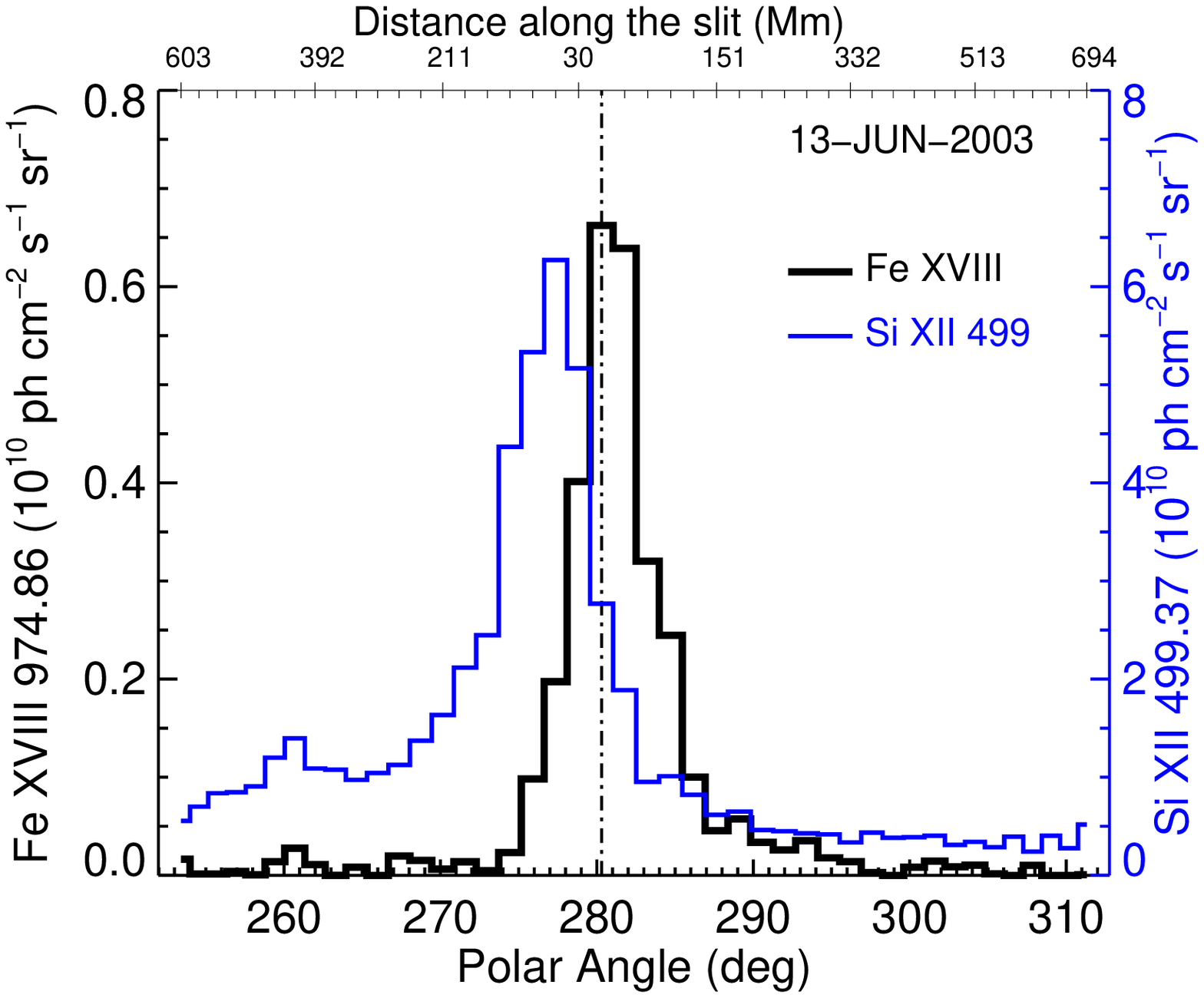} & 	
	\includegraphics[width=5.5cm]{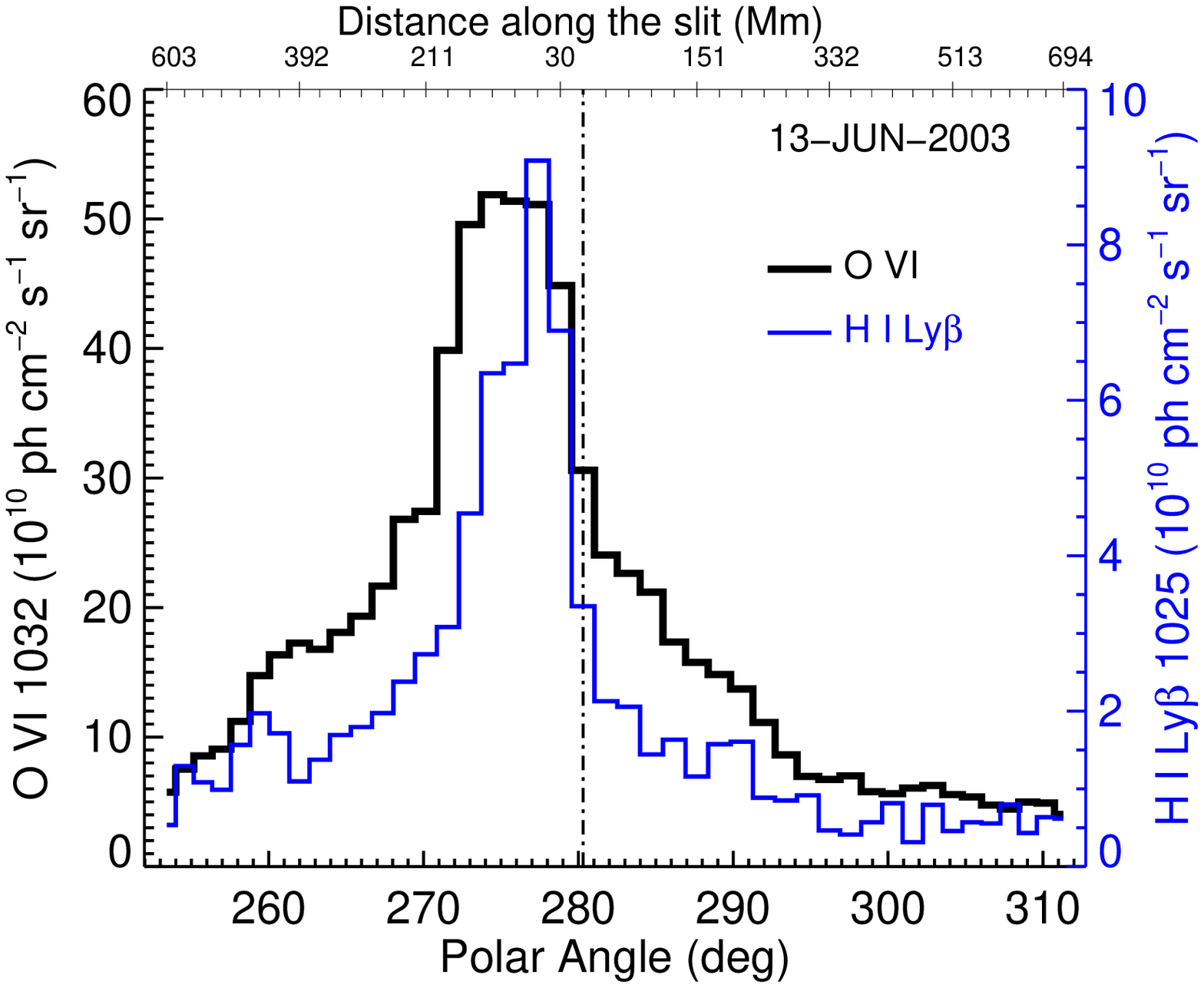} &
 	\includegraphics[width=5.5cm]{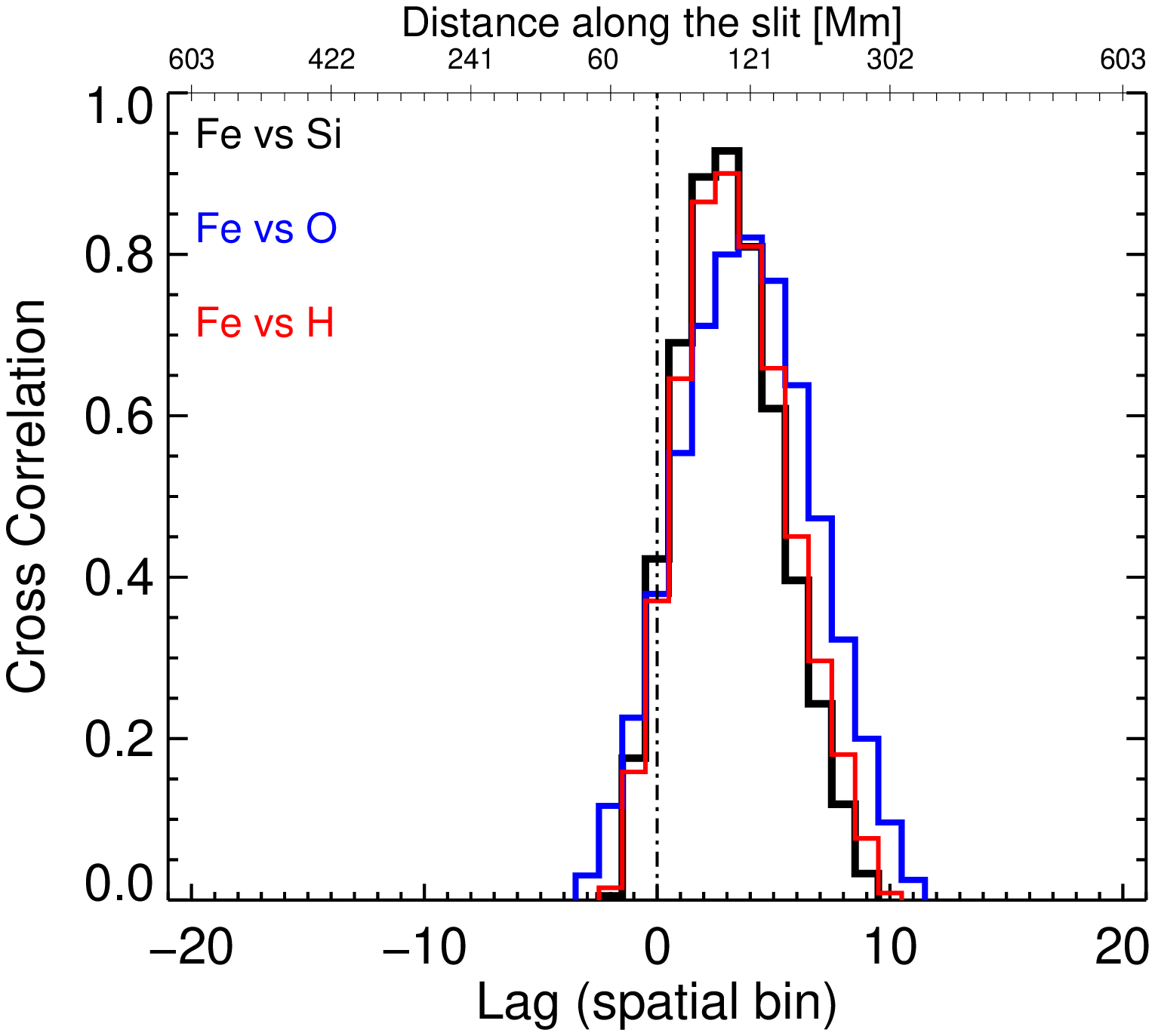} \\
 \end{tabular}
  \caption{The event of 2003 June 13 as an example of sharp [Fe~XVIII] and Si~XII that are significantly offset (class "O").  The panels present data as in Figure~\ref{fig1}. UVCS mirror is pointed to 1.67~\RSUN.}
\label{fig2}.
\end{figure*}

\begin{figure*}[h]
\center
   \begin{tabular}{ccc}
	\includegraphics[width=5.5cm]{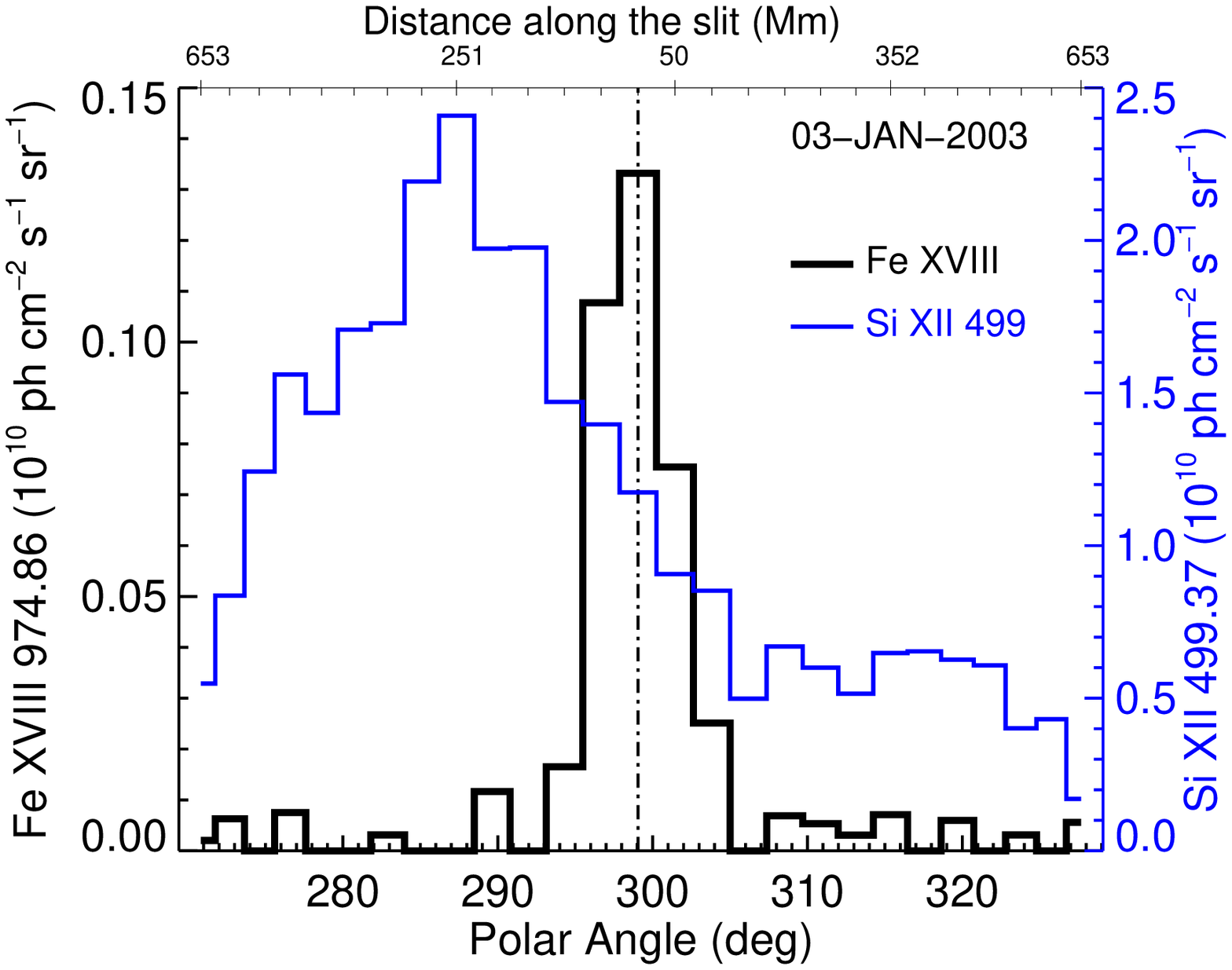} & 	
	\includegraphics[width=5.5cm]{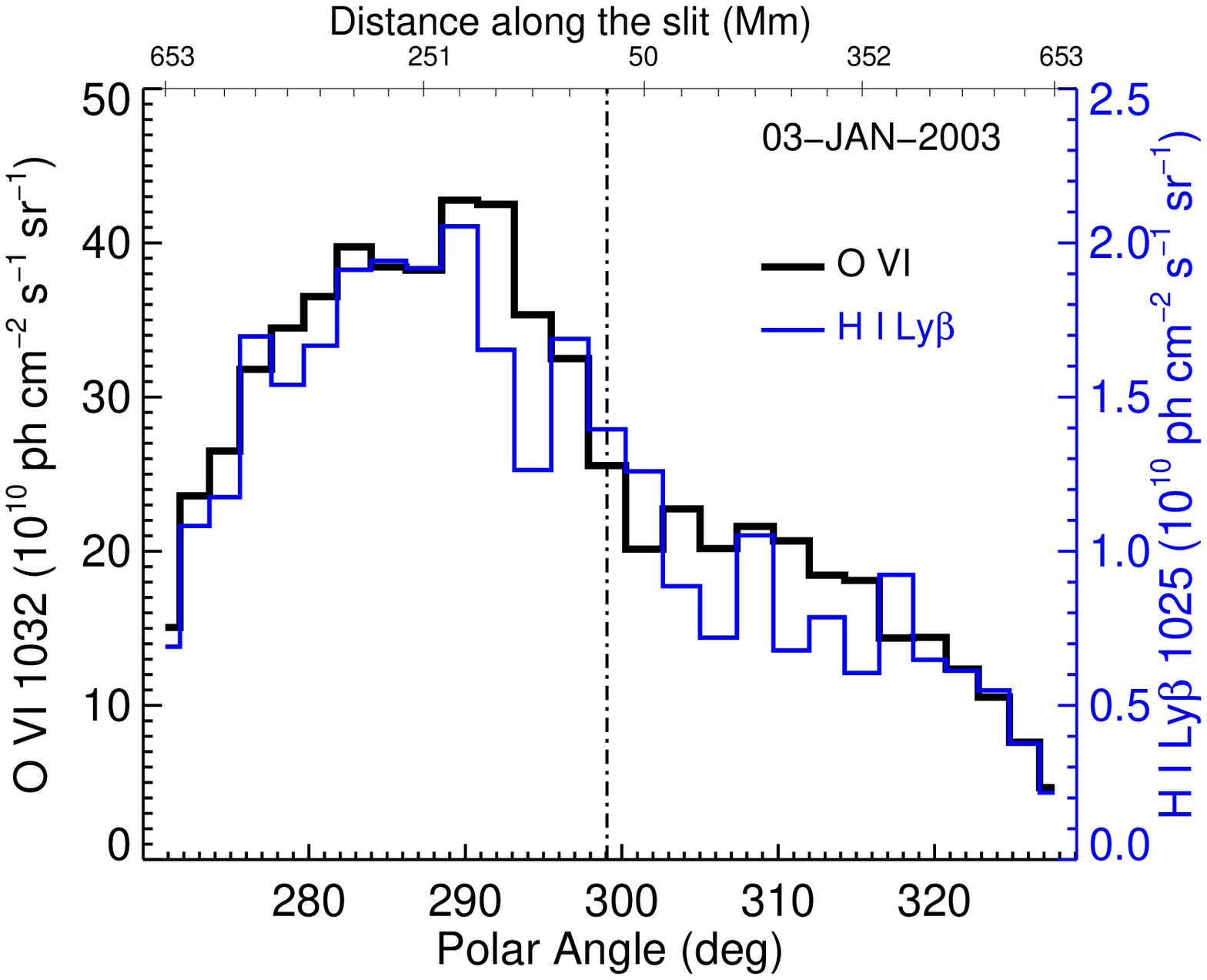} &
 	\includegraphics[width=5.5cm]{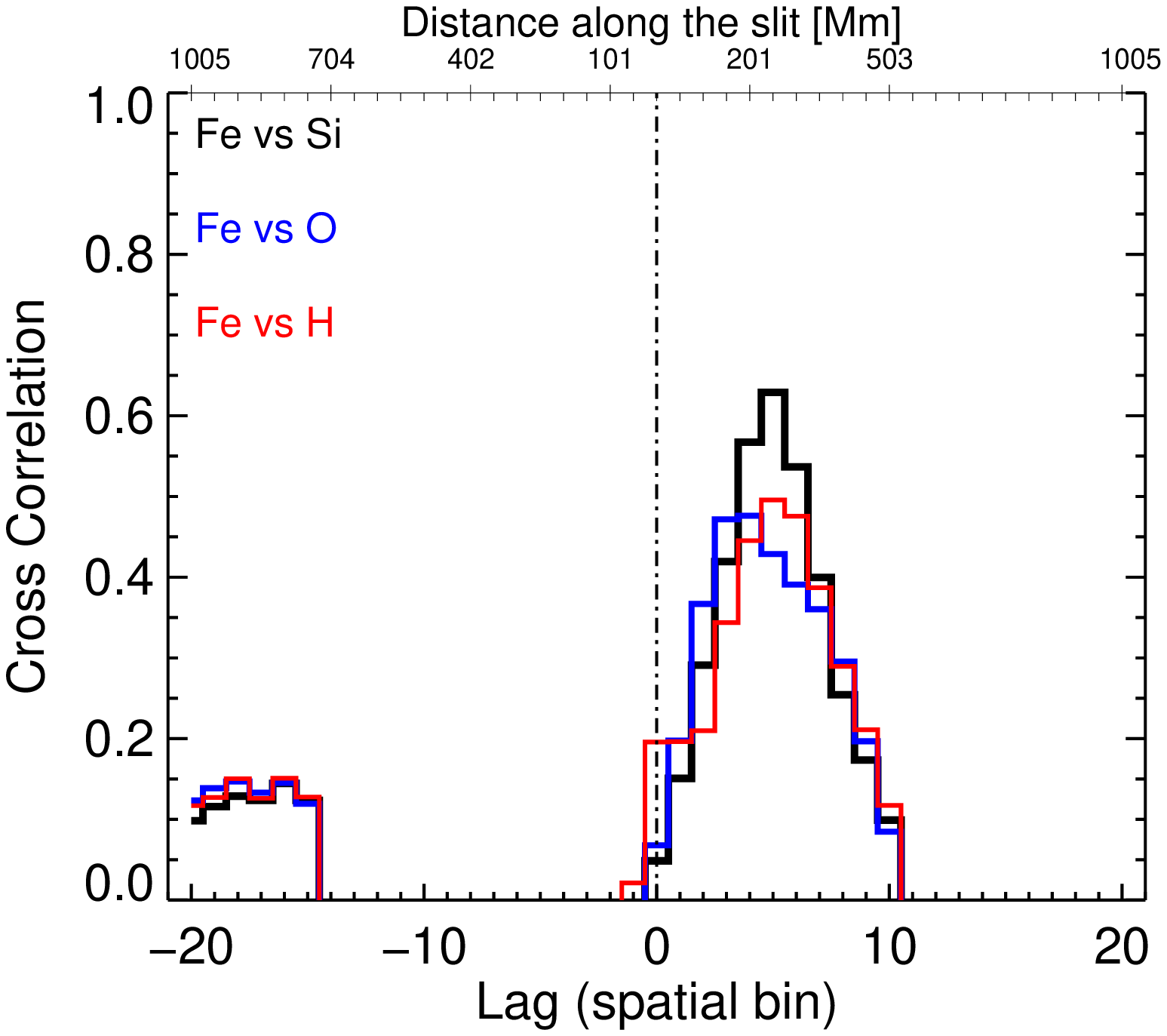} \\
 \end{tabular}
  \caption{The event of 2003 January 03 as an example of an event with no feature in Si~XII
that is correlated with the [Fe~XVIII] feature (class "O").  The panels present data as in Figure~\ref{fig1}. UVCS mirror is pointed to 1.72~\RSUN.}
\label{fig3}.
\end{figure*}

\begin{figure*}[h]
\center
   \begin{tabular}{ccc}
	\includegraphics[width=5.5cm]{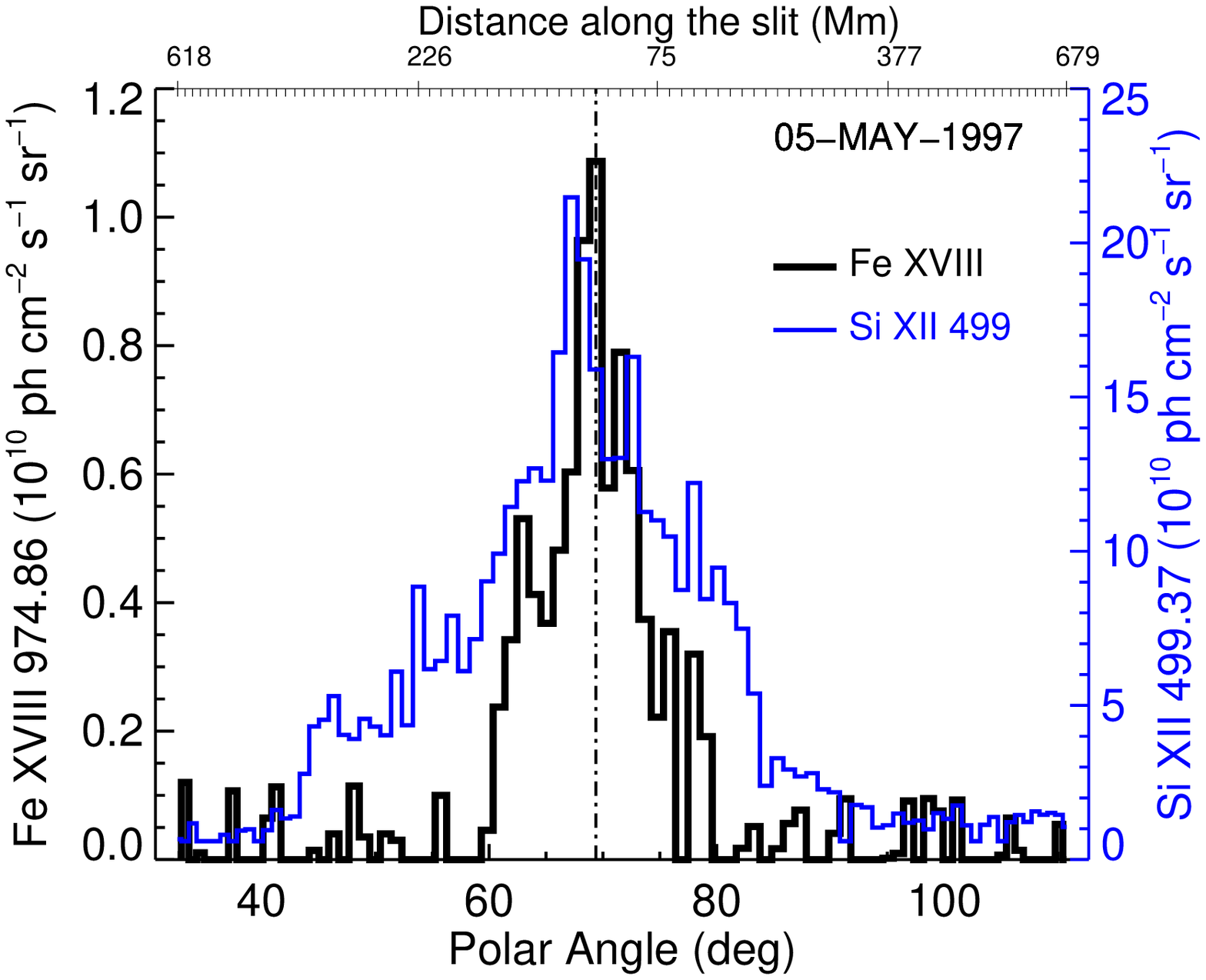} & 	
	\includegraphics[width=5.5cm]{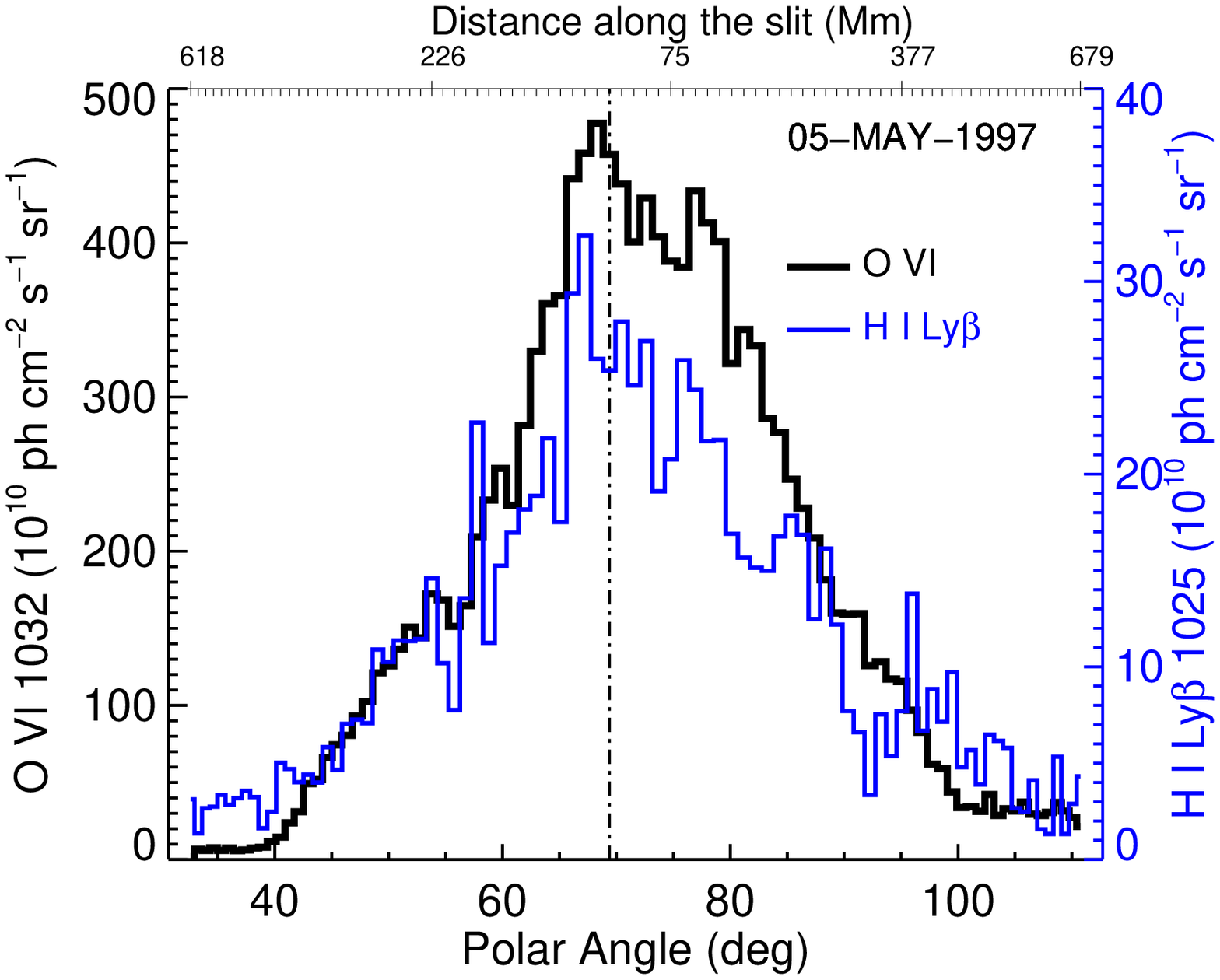} &
 	\includegraphics[width=5.5cm]{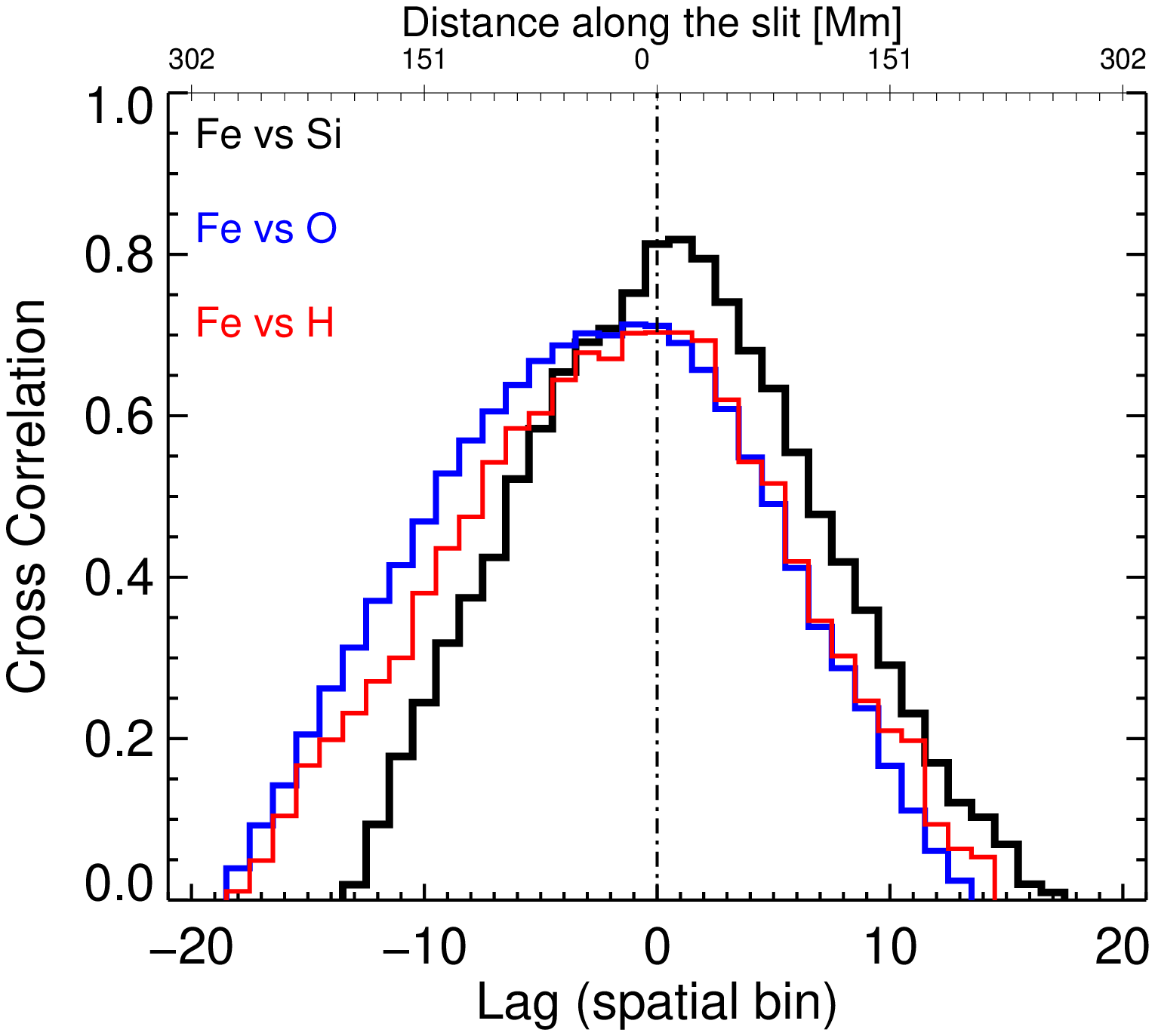} \\
 \end{tabular}
  \caption{The event of 1997 May 05 as an example of an [Fe~XVIII] feature within a
broader Si~XII emission feature (class "I").  The panels present data as in Figure~\ref{fig1}. UVCS mirror is pointed to 1.14~\RSUN.}
\label{fig4}.
\end{figure*}

We initially took 0.8 as the threshold for a meaningful cross correlation coefficient, but
in several cases and unrelated emission far from the CS reduced the correlation coefficient 
in spite of a close relationship in the vicinity of the CS.  Therefore, we relax the threshold
to 0.7 and list the coefficients, lags and widths in Table 1.   We note that in quite a few cases,
the Ly$\alpha$ and O~VI peaks are very broad and are probably unrelated foreground or background
emission, but in some cases there is a narrow peak of emission related to the current sheet
as well.   Figures~\ref{fig1} through
~\ref{fig4} present examples of 1) an event with [Fe~XVIII] and Si~XII essentially coincident,
2) an event with strong [Fe~XVIII] and Si~XII peaks offset from each other, 3) an event with
no meaningful correlation between [Fe~XVIII] and the other lines, and
4) an event with [Fe~XVIII] and Si~XII peaking at the same place, but with Si~XII spatially wider 
than [Fe~XVIII].

\begin{figure*}[]
\center
   \begin{tabular}{ccc}
	\includegraphics[width=5.5cm]{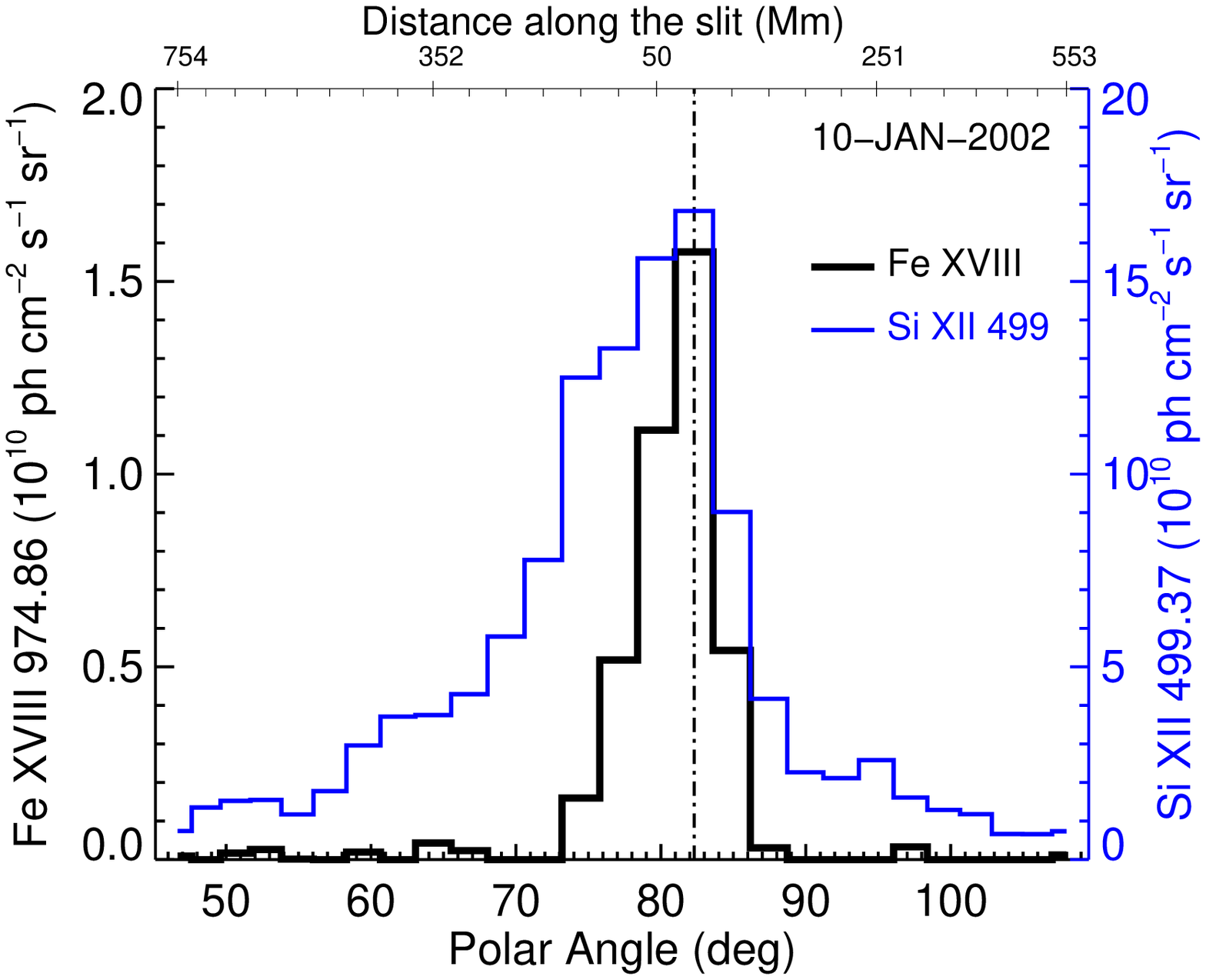} & 	
	\includegraphics[width=5.5cm]{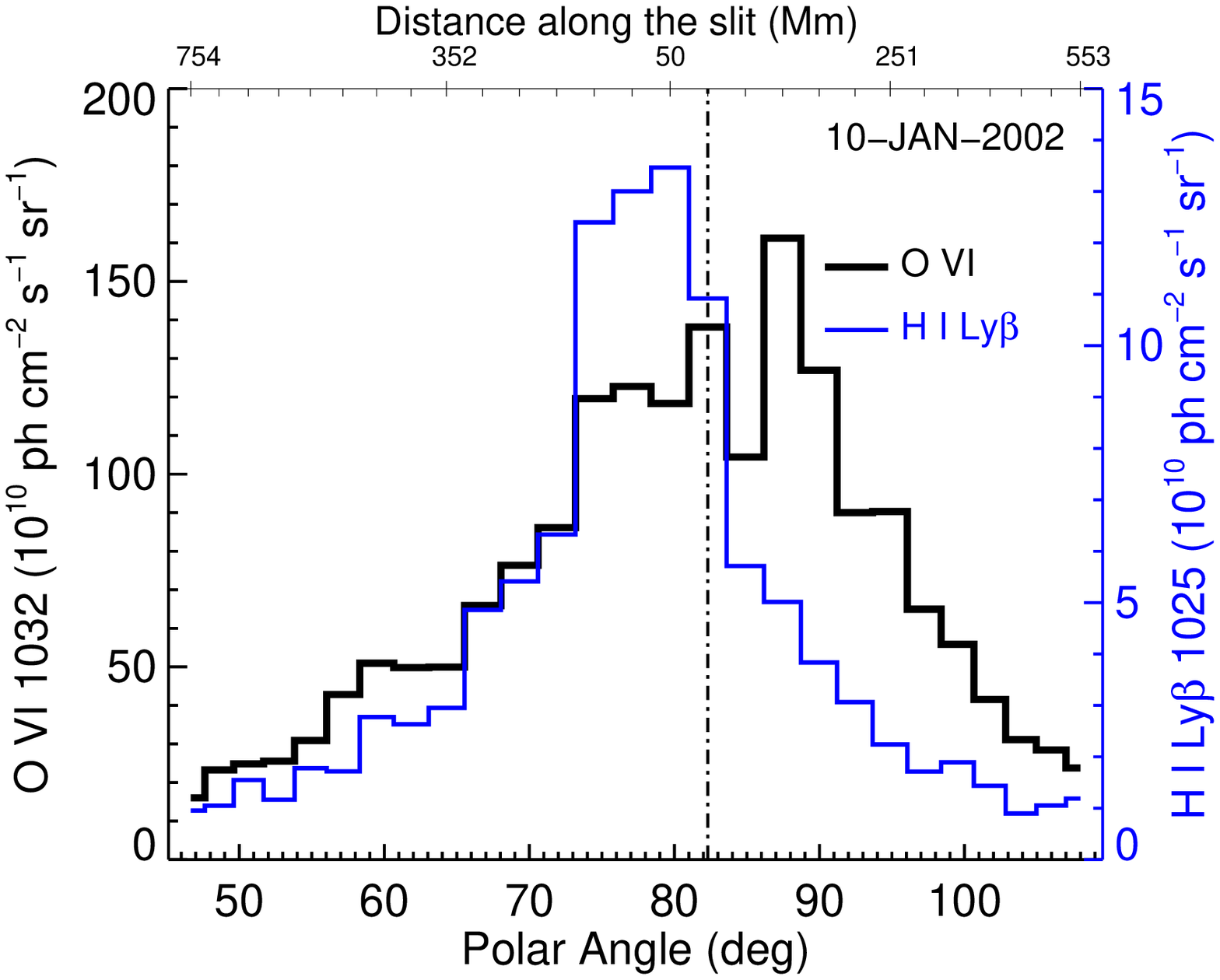} &
 	\includegraphics[width=5.5cm]{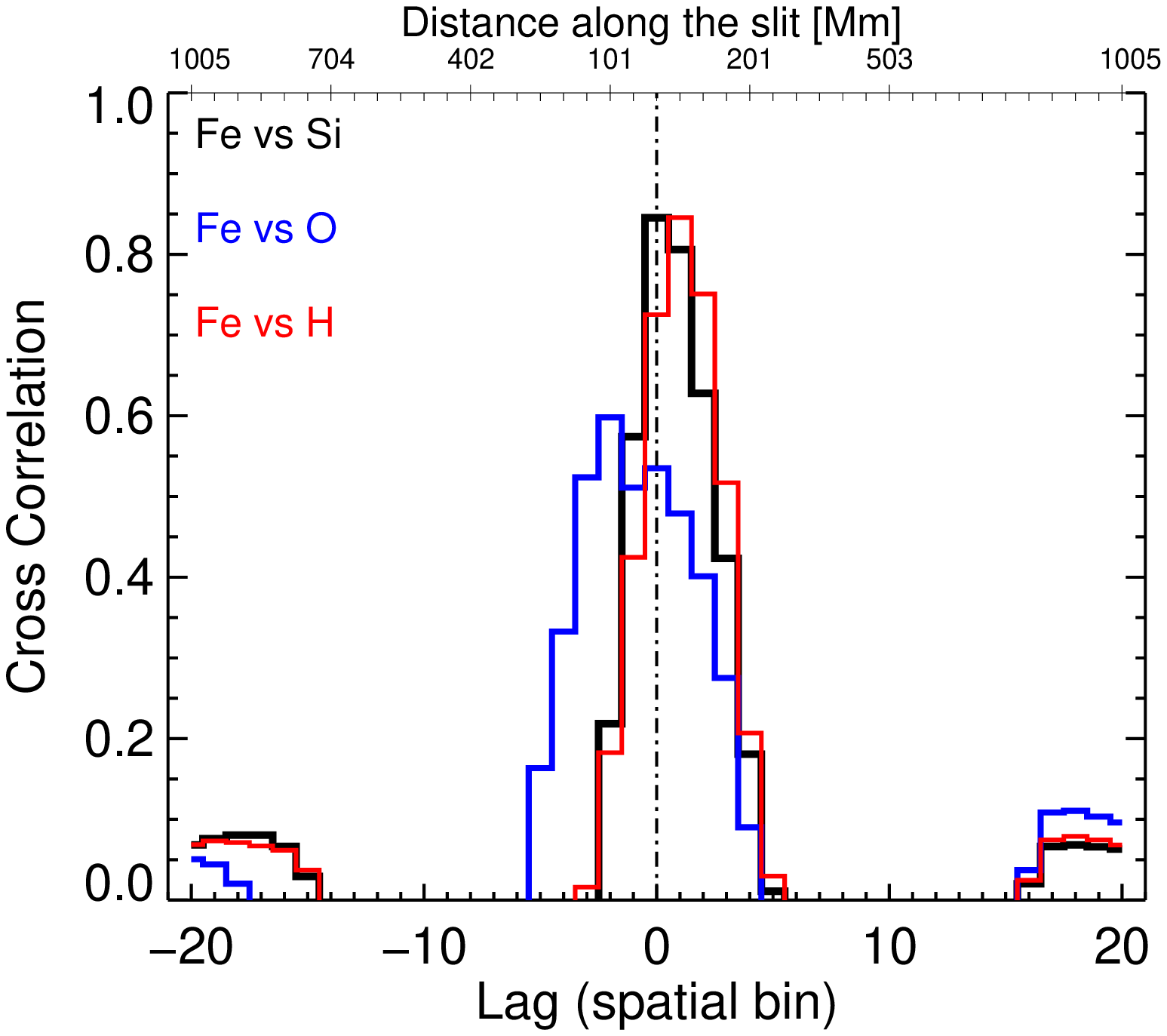} \\
 \end{tabular}
  \caption{The event of 2002 January 10 as another example of an [Fe~XVIII] feature within a broader Si~XII emission feature (class "I").  The panels present data as in Figure~\ref{fig1}. UVCS mirror is pointed to 1.57~\RSUN. In this case the [Fe XVIII] and Si XII
  peaks occur in the same place, but the [Fe XVIII] peak is symmetric and the Si XII peak is not.}
\label{fig5}.
\end{figure*}

\section{Interpretation}

\subsection{Observational signatures of physical structure}

In the simplest picture of a current sheet, the one evoked by 2D
cartoons of solar eruptions, cool material
flows symmetrically into the CS from the sides, is strongly heated and flows out
the top and bottom of the CS.  Because dissipation of magnetic flux decreases
the magnetic pressure, the density increases in the current sheet if the plasma
$\beta$ is low.  Thus one expects a thin sheet of [Fe~XVIII]
and Si~XII emission due to the high temperature and increased density
in the current sheet.  (For the purposes of this paper, "thin" means that the thickness
is small compared to the scale of the CME, but the CS can still be orders of magnitude
thicker than the classical CS thickness of the Sweet-Parker mechanism.)  There 
may also be a gap of the emission in the 
cooler lines, though a gap will only appear if the current sheet 
occupies a substantial fraction of the line of sight within a spatial bin.
The reduced ionization fractions of O VI and H I at the higher temperatures 
would be at least partly compensated by the higher density, but Doppler dimming
associated with the CS outflow would reduce the emission in O~VI and
Ly$\beta$ and severely reduce the Ly$\alpha$ emission \citep{lin05}.

Several variations are possible.  If the central temperature is well above 
the formation temperature of [Fe~XVIII], there could be two sheets of 
[Fe~XVIII] separated by region that would be dim in all the lines.  One possible
example will be discussed below.
Consideration of time-dependent ionization \citep{ko10, shen13} could 
alter the emission patterns in the CS, since the ionization
state lags behind the jump in the electron temperature.
The time-dependent ionization could mean that the inflowing gas is 
compressed and heated, emits in the lower ionization lines, and gradually 
ionizes toward [Fe~XVIII], giving a thin [Fe~XVIII] layer sandwiched 
between Si~XII sheets which in turn lie between O VI sheets. A thermal
conduction halo around the current sheet \citep{imada11} could have somewhat
similar effects. 

In this simple picture,
which should apply to either Petschek or turbulent \citep{lazarianvishniac}
reconnection, the inflow speed is of order 0.05 $V_A$ and the outflow
speed is of order $V_A$ according to both observations \citep{lin05, yokoyama} and
theory \citep{liu17}.  The high temperature emission seen by UVCS often lasts for a long time
and persists as the flare X-ray emission subsides
\citep{bemporad06, susino13}, but it may require only a low level of reconnection
relatively high (1.5 to 2 \RSUN ) in the corona.  
At very late times the structure might tend
toward that of an ordinary helmet streamer, with a thin current sheet having
a slow reconnection rate and slow solar wind flows on both sides, as in the
description of the heliospheric current sheet by \cite{wang00}.

However, in cases with narrow, well-defined emission features we sometimes
see an [Fe~XVIII] feature offset along the UVCS slit from a narrow Si~XII feature by
30 to 250 Mm.  There are several interesting possible explanations for
this asymmetry.  There is also, of course, the possibility that the [Fe~XVIII]
and Si~XII features are unrelated, which is plausible if the features are
very broad and poorly correlated, but not if both are narrow and much brighter than usually seen
at that height in the corona.  Since there are several of the latter cases,
we must consider physical explanations.

1)  The reconnection may be asymmetric in the sense that the density or
magnetic field is
higher on one side than the other.  This situation has been explored by
\cite{cassakshay}, \cite{murphy12} and \cite{murphy15}.  If the plasma $\beta$ is fairly low,
the pre-eruption magnetic field strength is comparable on the two sides, so the Alfv\'{e}n
speed is higher on the low density side.  The maximum temperature increase (if heating dominates over bulk kinetic energy and non-thermal particles) is given
by energy available to heat the plasma,

\begin{equation}
\Delta T = \frac{2}{3} \frac{B^2}{8 \pi n k_B} = \frac{1}{3} \frac{\mu}{k_B} V_A^2
\end{equation}

\noindent
where B is the reconnecting magnetic field corresponding to the magnetic free
energy, n is the total particle density, $k_B$ is the Boltzmann constant and
$\mu$ is the mean particle mass.  Here $V_A$ is the Alfv\'{e}n speed corresponding
to the reconnecting field, so the Alfv\'{e}n speed including the guide field
may be higher.  Thus the density should be smaller in the higher temperature [Fe~XVIII]
side than in the Si XII side by roughly the ratio of temperatures, or typically a factor of 
1.5 to 2 for modest  (factor of 2) density contrasts across the current sheet. 
The relative brightnesses of the Si~XII and [Fe~XVIII] lines
depend on the densities, the ionization states and on the sizes of the high and
low density regions.  This simple description assumes that all
the dissipated magnetic energy on the two sides goes into heat, while in fact
significant fractions may go bulk kinetic energy and non-thermal particles. 

2) In the picture where the reconnecting current sheet is located within a structure
like a helmet streamer, the Si~XII could come from the region outside the current
sheet, and there is no reason that the regions on the two sides should be the
same.  The densities, temperatures, flow speeds and thicknesses of the Si~XII layers
could be very different, in which case the center of the Si~XII emission will be 
offset from that of the [Fe~XVIII] in the CS. 

3) A thermal conduction halo, in which some of the heat of electrons within the current
sheet warms the inflowing gas outside the CS, appeared in the numerical simulations of
\cite{imada11}.  For a Peschek exhaust flow, this requires that thermal conduction carry
the energy across the shocks that bound the exhaust.  The electric fields in the shocks
should inhibit this transport of energy by electrons, so existing models might 
overestimate the heating.

\subsection{Projection effects}

The above interpretations apply to a perfectly edge-on viewing angle, but in general
the line of sight does not lie exactly along the CS, and the CS is not exactly
planar.  Projection effects can explain some observed morphologies, though in other
cases that explanation seems strained.  One of the most interesting results from the
UVCS observations of current sheets is the apparent thickness of order tens of Mm,
since the Sweet-Parker model predicts thicknesses orders of magnitude smaller.  The thicknesses
are compatible with the predictions of either turbulent reconnection \citep{lazarianvishniac}
or the exhaust regions of Petschek reconnection \citep{vrsnak09}, though the observed non-thermal
line broadening favors the former \citep{bemporad08, ciaravellaraymond}.  However, 
the observed thickness is just
an upper limit if the CS is extended along the LOS direction and the viewing angle 
differs from edge-on.  The actual thickness can be separated from projection effects
if another constraint is available.  In the best-studied case, that of 2003 November 4,
a white light observation from MLSO provides the electron column density, $N_e$, and
UVCS provides the emission measure, EM.  Since $N_e$ is the electron density times the
depth along the line of sight, and EM is density squared time depth along the line of
sight, one can divide one into the other to determine both
the density and the thickness of the current sheet \citep{ciaravellaraymond}.

In regards to offsets between Si~XII and [Fe~XVIII] emission, consider a current sheet near
the solar limb that is extended along the line of sight and whose temperature decreases from the near
end to the far end.  In that case, Si~XII will be brighter near the far end, and [Fe~XVIII]
will be brighter at the near end.  If the sheet is edge-on, the intensity peaks
will coincide, and the [Fe XVIII] to Si XII ratio will give an average temperature.
However, if the CS is tilted with respect to the LOS, the [Fe~XVIII] and Si~XII will 
appear to be offset. In the simple case of a linear change in temperature, 
the [Fe~XVIII] to Si~XII ratio and the temperature derived from that ratio will appear
to change continuously along the slit.  In the case of the 2003 June 13 observation
shown in Figure 2, a temperature changing from $2.5 \times 10^6$ to $5 \times 10^6$ K
across the Si and Fe features could account for the observations.  However, there is 
likely to be some foreground or background Si~XII emission, and temperatures away from
the peaks are not reliable.

There are two kinds of morphology that are difficult to explain with projection.  In a 
case where the CS appears as a deficit in low temperature emission lines, a thin current
sheet seen at an angle could produce only a very slight reduction of the brightness.  This
situation was reported by \cite{lin05}.  The second case is one where the [Fe~XVIII] is
located inside a broader Si~XII feature, as is seen in Figure~\ref{fig4}.  In principle, 
there could be a wavy sheet with temperature variations such that the hottest portions 
happen to lie at the projected center, while the cooler portions are off to the sides, 
but such a geometry is very contrived.  A CS extended along the LOS with a temperature
peaking in the middle is possible, but somewhat ad hoc.

Overall, projection effects can potentially explain simple offsets between Si~XII and
[Fe~XVIII] peaks if the CS is sufficiently extended along the LOS, hotter at one end
than the other, and seen at a modest inclination. It is difficult to distinguish between
this explanation and physical explanations such as asymmetric reconnection or a current
sheet embedded in an asymmetric helmet streamer structure.
 Stereoscopic observations might determine which explanation is correct.
Estimates of the length and orientation of the CS based on the magnetic configuration
before the eruption (for instance \cite{lee09}) could offer some less direct constraints.


\begin{deluxetable*}{llcccllll}
\tabletypesize{\footnotesize}
\tablecaption{\label{tab}}
\tablehead{ Event 	& UT 	& Height  	& PA		& Fe XVIII$^c$ 	& \multicolumn{3}{c}{Cross  Correlation Parameters}  & Class \\
                  		&    	&     		&		&      			& \multicolumn{3}{c}{(r)(lag - width) $\times 10^6$ m} & \\
                		&  	& \RSUN   	&  $^\circ$	&  $10^6$ m         &  Si XII  		& O VI 	& Ly$\beta$($\alpha$)  	&}
\startdata
97-05-05 &17:36 & 1.14      & 67          &  143  & (0.8) 15 -211    & (0.7) 0 -271     & (0.7) 30 - 256 & I \\
98-03-23$^b$ &16:00 & 1.51   & 235    &  164 & (0.8) 30-181    & (0.8) 0-241       & (0.7) 120-241 &  I\\
98-04-20 &17:10 & 1.20   &245         &  195 &(0.9) 0-181       & (0.9) 60- 211   & (0.9) -30 -181 & S\\
              & 17:46 & 1.25   &244            &   202 & (0.9) 0-241      & (0.9) 60-241    & (0.9) 30 - 241 & S\\
               &18:01 & 1.31     &244          &  199  & (0.8) 0-241      & (0.9) 60-241    & (0.8) 30-241 & S\\
               &18:39 & 1.36      &245         &  219  & (0.7) 0-271      & (0.8) 90-302    & (0.7) 120-332 &S \\
               &18:50 & 1.41     &245          &  225  & (0.8) 0-332      & (0.8) 60-302    & (0.8) 30-241 & S\\
               &19:36 & 1.46     &244          &  328  & (0.8) 0-302      & (0.8) 90-332    & (0.8) 30-302 & S \\
00-02-22 &18:39 & 1.45   &46      &  79    & (0.7) 151-302  & (0.7) 211-332  & (0.7) 241-392 &O \\
              & 18:50 & 1.51    &48          &  156  & (0.7) 180-362  & (0.7) 241-332  & (0.6) 241-392 & O \\  
01-10-31 & 04:27$^a$ & 1.53 &92       &  139  & (0.8) 166-196    & (0.6) 15-377      & (0.6) 150-347 &O \\
01-11-02 &02:23$^a$ & 1.52   &273      &  173  & (0.7) 0-211     & (0.5) 0- 362      & (0.6) 0-302  & S\\
01-12-20$^b$ &23:20 & 1.59$^a$ &271   &  125  & (0.7) 150-302 & (0.5) 301-402    & (0.6) 150-302 & O\\
            &         & 1.56  &281       &  174  & (0.8) 0-302     & (0.6) 100-402    & (0.8) 0-302  & S\\            
01-12-21     & 09:20  & 1.91 &280    &   464 & (0.9) 150-352  & (0.8) 251-402 & (0.8) 150-352 & S \\           
02-01-10 &20:45         & 1.57 &280        &  116  & (0.9) 0-201    & (0.6) 100-302  & (0.9) 50-201  &I\\
03-01-03 & 11:35 & 1.72 & 299 & 121  & (0.6) 251-201 & (0.5) 201-302 & (0.5) 251-251 & O \\
03-06-02    &    06:08 & 1.73$^a$  &250&  225  & (0.4) 60-814 & (0.6) 150-221 & (0.7) 90-151 &I \\
03-06-02      &  10:06 & 1.69       &257   &  124  &(0.7) 30-151 & (0.7) 30-181  & (0.7) 0-90 &I\\
03-06-13 &17:23 & 1.67   &280         &  113   & (0.9) 90-151 & (0.8) 121-211 & (0.9) 90-151 &O \\                
03-11-03$^b$& 11:07 & 1.75 &269         &  180   & (0.8) 30-181  & (0.7) 120-211 & (0.7) 90-181 & I\\
03-11-04 & 20:30 & 1.69  &253       &  70  & (0.9) 0-90  & (0.7) 181-211 & (0.8) 0-241  &S\\
\enddata
\tablenotetext{a}{low statistics}
\tablenotetext{b}{double peak }
\tablenotetext{c}{spatial size of the  current sheet in Mm, evaluated as the FWHM of the  Fe XVIII intensity distribution along the UVCS slit.}
\end{deluxetable*}


\section{Discussion and Summary}  

Magnetic reconnection in current sheets is an integral part of theoretical models
of solar flares and CMEs.  Structures observed in the UV have been interpreted
as examples of these current sheets and used to investigate properties of the
reconnection.  Some of these structures show a surprising offset between the
features seen in Si~XII emission lines and the [Fe~XVIII] line formed at a
temperature 2 to 3 times higher, with offsets of 30 to 180 Mm in
several events.  

The classification in Table 1 is somewhat subjective, but we see 6 events in which the structures
seen in Si~XII and [Fe~XVIII] are basically the same, suggesting a simple structure with a
constant temperature or range of temperatures.  It is also possible that these events
have current sheets with different temperatures at the front and back ends, but that the
CS is very closely aligned with the line of sight.   How close this alignment must be depends
on the length of the CS, which is not known except in the case of the 2003-11-04 event, where
the actual thickess and LOS length were comparable and the uncertainty in the offset precludes
an accurate determination of the angle or a reliable limit on the temperature change along
the CS.

Table 1 also shows 4 offset ("O") events.  The number of "O" events is similar to the number of 
"S" events with similar Si~XII and
[Fe~XVIII] structures.  That would be consistent with line-of-sight depths comparable to the the 
apparent thicknesses and a random distribution of viewing angles.  
Given the small number of events, projection of a CS seen not quite edge-on
seems capable of explaining the observations, though the symmetric [Fe~XVIII] peaks in
well resolved events such as 2003-06-13 (Fig. 2) would then be a coincidence.  The observations
could also be explained by asymmetric reconnection, which would require that the cooler and
hotter sides of the CS remain separate.  That question has not been adressed for turbulent CS,
and it would seem problematic in the classical Petschek picture, but a guide field might maintain
the temperature separation.

Four events show an [Fe~XVIII] peak lying within a broader Si~XII structure.  One plausible
interpretation is that the Si~XII is a helmet streamer, and the [Fe~XVIII] represents a hot, narrow
CS within it.  In the case of the 2002-01-10 event (Figure 5), the difference between the Si XII structure
and the structures seen in Ly$\beta$ and O VI, both of which show a drop where the [Fe~XVIII]  peaks,
suggests that the helmet streamer itself is asymmetric.  The intensity drop in the cool lines
indicates that the hot CS gas must occupy a substantial fraction of the volume near the
[Fe~XVIII] peak.  We note, however, that in many events the peak seen in [Fe~XVIII] and Si~XII
seems virtually unrelated to the much broader emission in O VI and the Lyman lines, and in
these cases the cool emission might arise in foreground or background gas unrelated to the CS.

Only one event, that of 2001-12-20, shows a double peak in [Fe~XVIII] that might have a physical
origin.  The others have a pre-existing peak, sometimes associated with an earlier CME as in the
1998-03-23, 2003-11-03 and 2003-11-04 events, and a new peak appears nearby after a CME.  That
also might be the explanation for the 2001-12-20 event, since the [Fe~XVIII] peak near PA=280$^\circ$ remains
constant while the narrow bright peak at PA=271$^\circ$ fades away.  However, a physical interpretation
in terms of gas hotter than the [Fe~XVIII] peak is also possible.  

Overall, offsets between the CS structures seen in [Fe~XVIII] and Si XII can have an interesting physical
interpretations such as asymmetric reconnection, or the mundane interpretation in terms of  projection
effects.  The currently existing data does not allow us to discriminate among them with certainty,
because a configuration can be invented to explain almost any observation with projection effects.
However, in some cases (simple offsets) the projection explanation is plausible because it arises from a 
configuration that is likely to occur, while in other cases (narrow [Fe~XVIII] centered in a broad Si XII feature) 
it would require a much less probable configuration.  Further
progress could come with higher quality observations that permit study of fine scale structure, temporal
variations or more detailed temperature and density determinations.  Other avenues for resolving the
ambiguity would be stereoscopic viewing or MHD modeling that could predict the morphology of the current 
sheet.  Perhaps a systematic study of CS seen during flares with AIA will reveal the extent of temperature 
variations along those structures and make the projection hypothesis more or less favored.

\acknowledgments
This work was partially supported by NASA grants NNH14AX61I and NNX13AG54G
and by N00173-14-1-G908 the NASA LWS grant NNH13ZDA001N.

\facility{SOHO (UVCS)}.

\appendix

\section{Appendix material}

{\it 1998-03/23:}  This event was studied by \cite{ciaravella02}.  There is a narrow, bright [Fe~XVIII]
peak, which we take to be the CS, and a weaker more diffuse peak to the south, which we take to be a long-lasting, bright active region.

{\it 2001-12-20:} This event shows a very clear double peak in [Fe~XVIII] with a 180 Mm
separation during the first 2 hours,
which evolves to a narrow peak offset from Si~XII and finally a broader peak coincident with
Si~XII.  The Si~XII structure is similar to that of Ly$\beta$, and it remains fairly constant.  

{\it 2002-01-10:} This long-lasting CS was studied by \cite{ko03}.  There is a dip in the low
temperature emission corresponding to a peak in the high temperature lines, indicating that
the CS is extended along the line of sight and that its thickness is comparable to the
spatial resolution of the observations.  

{\it 2003-06-02:} Details of this event are shown in \cite{schettino10} .

{\it 2003-06-13:} This event is unusual in that fairly narrow O VI and Ly$\beta$  show a similar structure
to Si~XII, somewhat offset from [Fe~XVIII].

{\it 2003-11-04:} This event was studied by \cite{ciaravellaraymond}.  For the 
analysis of this CS we used the data from 22:59/04 to 01:59/05  to exclude
several blobs of very cool, very bright emission in the H~I, C~III and O~VI lines
about 250-300$^\prime$$^\prime$, or about 180-240 Mm away from the [Fe~XVIII]
feature.  These blobs are probably small, secondary ejections of prominence material that
are channeled along field lines close to the CS, probably in a streamer-like structure.  
The correlated feature in Ly$\beta$ indicates a high density in the CS.

\end{document}